# Critical value of the total debt in view of the debts durations

**I.A. Molotkov, N.A. Ryabova**

N.V.Pushkov Institute of Terrestrial Magnetism, the Ionosphere and Radio Wave Propagation, Russian Academy of Sciences,
IZMIRAN, Moscow, Troitsk 142190, Kaluzhskoe shosse, 4, Russia
e-mail:  iamolotkov@yandex.ru, ryabova@izmiran.ru

Yanuary  29, 2016

**Abstract**

Parastatistic distribution of a total debt owed to a large number of creditors considered in relation to the duration of these debts. The process of debt calculation depends on the fractal dimension of economic system in which this process takes place. Two actual variants of these dimensions are investigated. Critical values for these variants are determined. These critical values represent the levels after that borrower bankruptcy occurs. The calculation of the critical value is performed by two independent methods: as the point where the entropy of the system reaches its maximum value, and as the point where the chemical potential is zero, which corresponds to the termination of payments on the debt. Both methods lead to the same critical value. When the velocity of money circulation decrease, it is found for what dimensions critical debt value is increased and for what it is decreased in the case when the velocity of money circulation is increased.

# 1 Introduction

The economic problem of the distribution of debts (loans) for their duration is discussed. The critical (threshold) value of the total debt is calculated. This value describes the boundary of qualitative change in the system defining a breach in economic security. The methods for calculating these critical values were developed by Maslov [1].

The parallels between statistical and thermodynamic properties of systems of physical particles and economic systems [2, 3] play an important role in this work. As in physics, the concept of entropy is also significant in economics. We are referring to Hartley's entropy [4], wherein all possible options for the debt calculation are equally probable.

Next, we deal with the values of individual debts $s_j$ ($j = 1, 2, ...k$). It is assumed that the number of debts $k$ is large. The value of each debt is characterized by its due date, or duration, $l$. The ratio $s/l$ is the main characteristic of the debt for the borrower (debtor); it determines the effort the borrower has to make to pay off the debt.

The distribution of debts is considered with the following limitations in mind: we will not take into account the interest and the derivatives incurred by the debts, nor consider past due ("toxic") debts.

The main purpose of this article is to calculate the value of a critical debt. At the end of this paper the use of long and short loans is analyzed.

# 2 The amount of debts, including durations

Let's say the debts have durations $l_j$. We will arrange the debts by duration, in ascending order, where $l_1 < l_2 < ... < l_k$, and consider the normalized reverse durations $l_k/l_j$ ($j = 1, 2, ...k$). Following the normalization and an arbitrary addition of virtual durations for which the value of debts $s_j = 0$, we receive a sequence of normalized reverse durations. Evidently, it is possible to assume such sequence to be integer:

$$\frac{l_k}{l_1} = k, \ldots, \frac{l_k}{l_j} = k+1-j, \ldots \frac{l_k}{l_k} = 1.$$

Let us consider the total debt. It is convenient to deal with dimensionless expressions that do not depend on the choice of monetary units. Let's introduce the value of the arithmetic mean of the debts ($\hat{s}$), and then consider each debt $s_j$ to be divided by the value of $\hat{s}$. Consider the dimensionless debt total:

$$\sigma = \sum_{j=1}^{k} s_j. \qquad (1)$$

Similarly, we introduce the dimensionless quantity of money

$$E = \sum_{j=1}^{k} \frac{l_k}{l_j} s_j = (k+1)\sigma - E_1, \quad E_1 = \sum_{j=1}^{k} j s_j, \qquad (2)$$

which is required to pay off all debts, taking into account their duration.

To determine the value of critical debt, it is necessary to choose the appropriate statistics. In physics, $k$ specifies the number of identical particles on a single quantum level. In contrast to Bose-Einstein statistics, where $k = \infty$, and Fermi-Dirac statistics, where $k = 1$, parastatistics is characterized by a finite value of $k$, which can be quite large and variable. As the number of creditors (and debts) $k$ during debt settlement can vary, the distribution of debts in relation to their durations should be considered a parastatistic distribution of identical elements (see [5]). The mathematical justification of parastatistic formulas was performed in [6].

When analyzing the physical systems it is important to indicate the fractal dimension of the system under consideration [7]. The economic system, in which debt calculation takes place, also has a certain dimension from which the velocity of money circulation is dependent. However, the calculation of the dimension of the economic system presents certain difficulties. Information on dimensions of the currency time series gives a certain ideas about dimensions of economic systems. The Hurst method [7, ch.8] permits to estimate dimensions of these series. For advanced countries with the stable currency the dimensions of currency series is found to be fractional with range in vicinity of $d = 1.5$ [8,9].

The case, when the dimension of economic system is equal to

$$d = 2, \qquad (3)$$

is the simplest for analysis. Examples of currency series show that variants of the fractional dimensions $1 < d < 2$ are of interest. Therefore we begin from detail examination of the case (3). Then we will consider the variant

$$1 < d \leq 2 - \delta, \quad 0 < \delta << 1. \qquad (4)$$

The investigation of variant (4) brings to results of debts calculation which essentially differs from results for the case (3). Another values $d > 2$ (including fractional $d$) are investigated analogously as in the case (3).

Thus, we suppose that $d = 2$. It follows from parastatistic distribution that [1]:

$$\sigma = \sum_{j=1}^{k} \left[ \frac{1}{\exp(b(j+\kappa))-1} - \frac{\sigma}{\exp(b\sigma(j+\kappa))-1} \right], \qquad (5)$$

$$E_1 = \sum_{j=1}^{k} j \left[ \frac{1}{\exp(b(j+\kappa))-1} - \frac{\sigma}{\exp(b\sigma(j+\kappa))-1} \right]. \qquad (6)$$

Here $b$ and $\kappa$ are positive parastatistic parameters defined by (5) and (6). Parameter $b$ refers to the inverse velocity of money circulation in the system (see [1, 3]). Parameter $\kappa$ takes into account possible changes in the number of the debts during settlement: in thermodynamics it corresponds to the chemical potential of the opposite sign. The volatility of the number of debts makes it necessary to consider the overall parastatistic distribution.

## 3 Asymptotic relations

Let us assume further that the quantities $\sigma$, $E$ and $k$ are large, and the debts $s_j$ are of the same order. Then, the constant $b$ will be small [1, 10]. Our next goal is deriving an asymptotic relationship between the quantities $E$, $\sigma$, $k$, $b$, and $\kappa$. The ultimate goal is to determine the critical values of $\sigma_0$ and $E_0$ for the total debt and for the total money required to pay it off over duration.

The equation we seek depends substantially on the multiplication results of $B = b\sigma$ and $bk$. It can be shown that with a small value of $B$, $\sigma_0$ is of no relevance. Therefore, we will further assume that this product is either greater than, or close to unity:

$$B \equiv b\sigma \geq 1 \tag{7}$$

Considering the behavior of the terms of series (5), we can see that with an increase $j$, the terms in the series decrease when $j + \kappa \geq 1/\sigma$ and do not contribute significantly to the value of the sum $\sigma$. This fact justifies replacing the difference

$$\exp\left[b(j+\kappa)\right] - 1 \tag{8}$$

by the first term $b(j+\kappa)$ of the series expansion. The condition (7) indicates that we consider periods of time during which money turnover in the system is of the same order as the value of the total debt, or lower.

The condition (7) and the approximation of the difference (8) give us the opportunity to transform the sums (5) and (6). As a result, we get:

$$\sigma = \frac{1}{b}\sum_{j=1}^{k} \frac{e^{(B(j+\kappa))} - B(j+\kappa) - 1}{\left[e^{(B(j+\kappa))} - 1\right](j+\kappa)}. \tag{9}$$

We will sum it up by means of the Euler – Maclaurin formula, and then divide both the numerator and the denominator of the integrand by $\exp\left[B(x+\kappa)\right]$:

$$\sigma \cong \frac{1}{b}\int_{1}^{k} \frac{e^{B(x+\kappa)} - B(x+\kappa) - 1}{\left[e^{B(x+\kappa)} - 1\right](x+\kappa)} dx =$$

$$= \frac{1}{b}\int_{1}^{k} \frac{\left[1 - B(x+\kappa)e^{-B(x+\kappa)} - e^{-B(x+\kappa)}\right]}{1 - e^{-B(x+\kappa)}} \frac{dx}{x+\kappa}. \tag{10}$$

Considering that, in view of (7), the value of $\exp[-B(x+\kappa)]$ is small compared to 1, we can get rid (10) of the denominator:

$$\frac{1}{1-\exp[-B(x+\kappa)]} \simeq 1+\exp[-B(x+\kappa)]. \qquad (11)$$

After deploying (11) and replacing the sum $x+\kappa$ with a new variable of integration, we obtain:

$$\sigma \simeq \frac{1}{b}\int_{\kappa}^{k+\kappa}\left[1-Bx\,e^{-Bx}-Bx\,e^{-2Bx}\right]\frac{dx}{x} \simeq \frac{1}{b}\left[\ln\left(\frac{k}{\kappa}+1\right)-e^{-B\kappa}\right]. \qquad (12)$$

Similarly to (12):

$$E_1 \simeq \frac{1}{b}\int_{\kappa}^{k+\kappa}\left[1-Bx\,e^{-Bx}-Bx\,e^{-2Bx}\right]\frac{(x-\kappa)}{x}dx \simeq$$

$$\simeq \frac{1}{b}\int_{\kappa}^{k+\kappa}\left[1-Bx\,e^{-Bx}\right]dx - \kappa\sigma. \qquad (13)$$

The evaluation of integrals in leading-order terms yields

$$E_1 = \frac{1}{b}\left[k-\kappa e^{-B\kappa}-\frac{1}{B}e^{-B\kappa}\right]-\kappa\sigma.$$

Together with equations (2) and (12), the last expression determines the amount necessary to repay the debts:

$$E = \frac{1}{b}\left\{(k+\kappa+1)\left[\ln\left(\frac{k}{\kappa}+1\right)-e^{-B\kappa}\right]-k+\kappa e^{-B\kappa}\right\}. \qquad (14)$$

Equation (12) in leading-order terms yields the equation

$$B \simeq \ln\left(\frac{k}{\kappa}+1\right)-e^{-B\kappa} \qquad (15)$$

between $B$, $k$ и $\kappa$, where the second term on the right serves for corrective purposes. Equation (15) simplifies the expression for the required amount:

$$E = \frac{1}{b}\left[(k+\kappa+1)B - k + \kappa e^{-B\kappa}\right]. \tag{16}$$

## 4 Entropy for *d*=2

Like in physics, the entropy in economics is defined by a logarithm of the number of possible implementation options in the system under study. Let's proceed to the computation of the entropy $S$. The point of maximum entropy viewed as a function of $\sigma$ determines the critical value of $\sigma_0$, see [1]. The calculation of entropy can be performed by taking into account the properties of parastatistic distribution, as it was done in [10]. Instead of this cumbersome way, we will use the analogy of the physical and economic variables and determine $S$ by means of the formula

$$S = K \frac{\partial E}{\partial (1/b)}, \tag{17}$$

which was established in [10, 11] for physical systems with fractal dimension $d \geq 2$. Here, $K$ is the proportionality factor, which depends only on the system's dimension. It is shown for the dimension value $d = 2$ that

$$K = 2\sqrt{6}, \tag{18}$$

see [10]. Another independent method of calculating $\sigma_0$ is presented in the next section. The coincidence of these two calculations justifies the use of formula (17) in econophysics and thus provides an additional confirmation for the analogy above.

Let's turn to calculating the entropy on the basis of equations (16) and (17). We determine the derivative $\partial E/\partial(1/b) = \partial E/\partial(V)$ ($V = b^{-1}$ - velocity of money circulation in the system). This derivative is calculated for fixed $\sigma$ and $k$, accounting for the dependence of $\kappa$ on $V$. This dependence is determined by (15), from which we can obtain that

$$\kappa'(V) \simeq \frac{\kappa}{V}\ln\frac{k}{\kappa} \cong \frac{B\kappa}{V}. \tag{19}$$

Let us rewrite (16) in this form:

$$E = (k + \kappa + 1)\sigma - kV + \kappa V \exp\left(-\frac{\sigma\kappa}{V}\right). \tag{20}$$

Differentiating (20) with respect to $V$ in view of the dependence $\kappa(V)$, we find that

$$\frac{\partial E}{\partial V} = -k + B^2\kappa + e^{-B\kappa}\left(\kappa + B\kappa - B^2\kappa^2 + B\kappa^2\right). \tag{21}$$

As a result, according to (17), (18), and after discarding lower-order terms, we obtain the following expression for the entropy

$$S = 2\sqrt{6}\,\frac{\partial E}{\partial V} \simeq 2\sqrt{6}\left[-k + kB^2 e^{-B} - e^{-B\kappa}B^2\kappa^2\right]. \tag{22}$$

The last step of calculating the critical value of $\sigma_0$ is setting to zero the derivative $\partial S/\partial \sigma$ (or, equivalently, $\partial S/\partial B$). The differentiation is performed for a fixed $k$, but with consideration for the dependence of $\kappa$ on $B$, $\kappa'(B) = -\kappa$. Assuming the derivative $\partial S/\partial B$ equals zero results in

$$B\kappa^2 = \kappa^2 + 2\kappa - e^{B\kappa}.$$

Transforming this relation by means of equation (15) in the higher-order, we find that

$$b\sigma_0 = \ln k - \frac{1}{k}. \tag{23}$$

Thus, the maximum entropy $S$ is reached at

$$\sigma = \sigma_0 \simeq \frac{\ln k}{b} = V \ln k. \tag{24}$$

Here $V$ means the velocity of money circulation in the system. If the velocity of money circulation is unknown, then by using the formulas (12) and (14) one can express the critical value of $\sigma_0$ via debt parameters of $E$ and $\sigma$:

$$\sigma_0 = (E - k\sigma)\frac{\ln k}{k}. \tag{25}$$

Equation (25) allows to express the velocity of money circulation in the system in an explicit form:

$$V = \frac{1}{b} = \frac{E - k\sigma}{k}. \tag{26}$$

## 5 Another method to calculate the critical value of $\sigma_0$

The idea of another method of calculating the critical value of $\sigma_0$ is suggested in [1]. The critical value of the total debt corresponds to a moment of debt settlement, in which the number of debts is no longer changing. At this point of calculation the value of $\kappa$ equals zero.

Taking into account expression (12), one can determine that

$$\sigma|_{\kappa=0} = \frac{1}{b}\int_1^k \left[1 - Bx e^{-Bx} + O\left(e^{-2Bx}\right)\right] \frac{dx}{x}, \tag{27}$$

where $B = B_0 = b\sigma_0$. By means of (27) and (15) we obtain

$$\sigma_0 = \frac{1}{b}\left[\ln k - \frac{1}{k} + O\left(\frac{1}{k^2}\right)\right], \tag{28}$$

which is consistent with formula (23). The matching results obtained from formulas (23) and (28) justify both the use of formula (17) and the applicability of the concept of entropy as a whole in economic problems. In (28), parameter $b = V^{-1}$ is small, but it does not depend on a large number $k$ - the number of creditors.

Thus, the value of the critical debt in the case of dimension (3) for a large number of creditors is determined by the expressions (23) or (28). Formulas (23) and (28) show that when $\sigma > \sigma_0$, the borrower is forced to declare bankruptcy, but when $\sigma < \sigma_0$, the borrower is able to repay the debt.

## 6 Changes in a debt's parameters during the repayment process

Let's consider the debt parameters at a critical point. All parameters at that point are tagged with a zero in formulae below. It has been already noted that the critical debt total $\sigma_0$ is determined via formulas (23) and (28). It follows from these formulas that the critical number of debts $k_0$ is determined by

$$\ln k_0 = \frac{\sigma_0}{V_0}. \tag{29}$$

The critical value of the chemical potential is

$$\kappa_0 = 0. \tag{30}$$

At the critical point equation (15) transforms into the following equation:

$$\sigma_0 = V_0 \left(\ln k_0 - \ln \kappa_0\right) - V_0 \exp\left(-\frac{\kappa_0 \sigma_0}{V_0}\right). \tag{31}$$

A joint consideration of (29) and (31) shows that these equations can be satisfied only with simultaneous approach to zero at the critical point of the parameters $V$ and $1/\ln k$. We determine that when approaching the critical point, the three values $\kappa$, $V$, and $1/\ln k$ tend toward zero. Equality $V_0 = 0$ means that money circulation related to paying off the debts stops at the critical point, which is consistent with the equality (30).

As a result, in the process of repayment, the values $V$ and $\kappa$ decrease to zero, and the debt total and the number of creditors decrease to $\sigma_0$ and $k_0$, respectively.

## 7 Critical value of the total debt for dimension (4)

Now, we start to investigate of the total debt dynamics in the case of dimension, which satisfies to condition (4). The parastatistic equation instead of (5) now has following form

$$\sigma = \alpha \sum_{j=1}^{k} \frac{1}{j^{1-\alpha}} \left[ \frac{1}{e^{b(j+k)} - 1} - \frac{\sigma}{e^{b\sigma(j+k)} - 1} \right], \tag{32}$$

where $\alpha = d/2$.

We performed a set of transformations which are similar to transformations at the case $d = 2$. With neglect of small terms we have:

$$\sigma = \frac{\alpha}{b} \int_{1+\kappa}^{k+\kappa} \frac{1}{x^{1-\alpha}(x+\kappa)} \left[1 - B(x+\kappa)\exp(-B(x+\kappa))\right] dx,$$

and as before $B \equiv b\sigma \geq 1$.

As above, to determine critical debt $\sigma_0$ we write the value $\sigma$ for $\kappa = 0$ in the form:

$$\sigma_0 = \frac{\alpha}{b} \int_1^k \frac{1}{x^{2-\alpha}} \left(1 - B_0 x \exp(-B_0 x)\right) dx = \frac{\alpha}{b}(\sigma_{01} + \sigma_{02}), \quad (33)$$

$B_0 = b\sigma_0$. The principal term in (33) is equal

$$\sigma_{01} = \frac{1}{1-\alpha}\left(1 - \frac{1}{k^{1-\alpha}}\right).$$

The correction term in leading order is

$$\sigma_{02} \cong -\frac{1}{2}\exp(-B_0) = \frac{1}{2}\exp\left(-\frac{\alpha}{1-\alpha}\right)$$

and

$$\sigma_0 = \frac{\alpha}{1-\alpha} \frac{1}{b}\left[1 + O\left(\frac{1}{k^{1-\alpha}}\right) + O\left[\exp\left(-\frac{\alpha}{1-\alpha}\right)\right]\right].$$

As a result, instead of formula (24) we have:

.
$$\sigma_0 \cong \frac{\alpha}{1-\alpha} V \quad (34)$$

The factors in the right hand of (34) are not independent. The decreasing of the dimension corresponds to the increasing of the velocity $V$ [12]. The quantities $\alpha$ and $V$ are connected by the low of energy conservation

$$E = f(\alpha) V^{1+\alpha}, \quad (35)$$

where function $f(\alpha) = \alpha^2 \Gamma(\alpha) \zeta(1+\alpha)$ ($\Gamma$ is the gamma-function, $\zeta$ is the Riemann zeta function) monotonically increases for all $\alpha < 1.7$ [10,12]. It is follows from here that the factor $V^{1+\alpha}$ monotonically decreases in spite of index $1+\alpha$ growth. The function $V(\alpha)$ is decreasing even more. Meanwhile the factor $\dfrac{\alpha}{1-\alpha}$ in (34) within the interval (4) is quickly increasing (from 1 to ∞). Therefore expression $\sigma_0$ in (34) grows together with $\alpha$, but is diminishing together with $V$.

Thus, the dependence of $\sigma_0$ on $V$ in the case (4) is radically different from this dependence in the case (3). Simultaneously the dependence of $\sigma_0$ on the number of creditors $k$ is found to be inessential in the case (4).

## 8 Short and long debts

Suppose we have $m$ short debts $s_1$ with the duration $L_1$ and $n$ long debts $s_2$ with the duration $L_2 \gg L_1$. Obviously, $m+n=k$, $k\hat{s} = ms_1 + ns_2$. As before, we assume the debts $s_1$ and $s_2$ divided by $\hat{s}$. Now, the dimensionless debt total is:

$$\sigma = m s_1 + n s_2, \quad (36)$$

and the dimensionless total payoff value

$$E = m s_1 \frac{L_2}{L_1} + n s_2. \quad (37)$$

We assume that the values of the debts $s_1$ and $s_2$ are of the same order and the numbers $m$ and $n$ are finite and are also of the same order. On the basis of formulas (23) and (28) one can determine that the critical debt is

$$\sigma_0 = V \ln(m+n). \quad (38)$$

Let's consider the cases of a dominance either in long or short-term debts (loans). The case $n \gg m$ does not yield fundamentally new results: the general formulas (28) - (31) are applicable.

More interesting is the case $m \gg n$ of the predominance of short-term debts. By using (26), (36) and (37) one can find that

$$V \simeq s_1 \frac{L_2}{L_1}, \qquad (39)$$

$$\sigma_0 \simeq s_1 \frac{L_2}{L_1} \ln m. \qquad (40)$$

We can see that in this case the turnover rate is high; also high is the critical value of $\sigma_0$, which in turn means an increase in borrowing capacity. While the benefits of this case are obvious, the formula (40) provides a quantitative estimate of these benefits.

# 7  Conclusion

The critical values $\sigma_0$ of the total debt are calculated for a large number of creditors and in view of debts duration. Thus, it is obtained the threshold after that bankrupt of borrower occurs. The conducted analysis assumes that totality of debts is subordinated to parastatistic distribution.

The form of the parastatistic relations depends on the fractal dimension of economic system in which debt calculation take place. Two most actual cases of dimensions are investigated: case (3) and case (4).

The necessity of the separate study of these cases is determined by the fact that boundary dimension d = 2 is a singular $\lambda$-point for parastatistic system [5,10,13]. This singularity determines specific behavior of the most important thermodynamic characteristic: the entropy in $\lambda$-point has point of inflection, the second derivative of chemical potential with respect to the velocity of money circulation (to the temperature) has jump in this point.

The critical values calculation are realized here by two ways: as the point of the entropy maximum value and as the point zero for the chemical potential. Both ways give one and the same critical value.

The dimension variants (3) and (4) lead to essentially different results. In the case (3) formulas (23) or (28) are obtained for critical values. In this case critical measurement $\sigma_0$ linearly depend on the velocity of money circulation in the system (which identically connected by the law (35) with dimension) and logarithmically related with the number of creditors. In the case (4) formula (34) is obtained. Taking in an account relation (35) it established that value $\sigma_0$ increase with $\alpha$ and decrease with $V$. The dependence on the number of creditors is inessential.

The coincidence of formulas (23) and (28), obtained by independent methods, proves the use the concept of entropy and chemical potential in economic problem.

Also it is shown that the use of short loans permits to increase the critical value $\sigma_0$ and thus gain borrowing capacity.